\documentclass[sigconf]{acmart}
\usepackage{float}
\usepackage[section]{placeins}
\usepackage[justification=centering]{caption}

\def\BibTeX{{\rm B\kern-.05em{\sc i\kern-.025em b}\kern-.08emT\kern-.1667em\lower.7ex\hbox{E}\kern-.125emX}}
    
\copyrightyear{2019}
\acmYear{2019}
\setcopyright{acmlicensed}
\acmConference[KDD '19]{KDD '19: ACM SIGKDD Conference on Knowledge Discovery and Data Mining}{August 04--08, 2019}{Anchorage, Alaska}
\acmBooktitle{KDD '19: ACM SIGKDD Conference on Knowledge Discovery and Data Mining, August 04--08, 2019, Anchorage, Alaska}

\begin{document}

%
\title{Large Scale Product Categorization using Structured and Unstructured Attributes}

%

\author{Abhinandan Krishnan}
\email{akrishnan@walmartlabs.com}
\affiliation{%
 \institution{WalmartLabs}
 \country{USA}}
 
\author{Abilash Amarthaluri}
\email{aamarth@walmartlabs.com}
\affiliation{%
  \institution{WalmartLabs}
  \country{USA}}

%
\renewcommand{\shortauthors}{Abhinandan and Abilash}

%
\begin{abstract}
Product categorization using text data for eCommerce is a very challenging extreme classification problem with several thousands of classes and several millions of products to classify. Even though multi-class text classification is a well studied problem both in academia and industry, most approaches either deal with treating product content as a single pile of text, or only consider a few product attributes for modelling purposes. Given the variety of products sold on popular eCommerce platforms, it is hard to consider all available product attributes as part of the modeling exercise, considering that products possess their own unique set of attributes based on category. In this paper, we compare hierarchical models to flat models and show that in specific cases, flat models perform better. We explore two Deep Learning based models that extract features from individual pieces of unstructured data from each product and then combine them to create a product signature. We also propose a novel idea of using structured attributes and their values together in an unstructured fashion along with convolutional filters such that the ordering of the attributes and the differing attributes by product categories no longer becomes a modelling challenge. This approach is also more robust to the presence of faulty product attribute names and values and can elegantly generalize to use both closed list and open list attributes.
\end{abstract}

%
%


%
\keywords{neural networks, text classification, extreme classification, eCommerce, structured datasets}

%
\maketitle

\section{Introduction}
Categorizing products into a hierarchical taxonomy has become a central part of the organizational efforts of eCommerce companies. It is a critical step that acts as a precursor to multiple downstream systems like search, facets etc. In the eCommerce context, it is posed as an extreme multi-class classification problem and is relatively well studied in both academia and industry. 

There are several challenges that are unique to product categorization in the eCommerce domain that are not observed in traditional extreme classification challenges like ImageNet. Product catalogs are in constant flux with old products being retired and new products being added on a daily basis. Due to the dynamic nature of the product catalog, the hierarchical taxonomy against which they are categorized is also in constant flux although not at the same rate as the items in the catalog. Hence, we need to classify products with acceptable performance against a non-stationary dataset where both the sample space and the label space change at the same time. This also means that both the training and validation sets need to keep changing to reflect the latest snapshot of the distribution of the catalog. Acquiring labeled data for this changing catalog is also an extremely expensive process and hence intelligent sampling strategies need to be employed to reuse as many previously labeled examples as possible. With marketplace platforms such as Amazon, Walmart, eBay etc., there are new sellers and vendors being added everyday which results in a wide distribution of data quality levels for the products being setup. While most products have some common attributes like title, description, image etc., every product also has a unique set of structured attributes describing it depending on the category the product belongs to. The total set of unique attributes in the catalog is in the tens of thousands (N) while an individual product might only possess a few attributes (k<<N) that are relevant to it. In addition, the quality of product attributes widely varies by seller. Each product attribute also has its own value space that presents a modelling challenge. All these added complexities make product categorization an extremely challenging problem to tackle.

Next, in section 2, we briefly review related work in extreme multi-class classification and product categorization in particular. Section 3 outlines our preliminary approach to classification using hierarchical multi-class models, our move to a flat classification scheme using two different flavors of Deep Learning models, a baseline architecture for structured attributes using word averaging and finally an innovative way to use all available structured attributes in a product in an unstructured format along with convolutional filters. In section 4, the experimental setup is described, specifically the details about the dataset, preprocessing, training schedule, dictionaries, embeddings, models and their deployment. Finally in section 5, we provide some results comparing the different approaches we have experimented with.

\section{Related Work}
Many websites, especially in eCommerce, organize their product catalog into concept hierarchies or taxonomies. Some common examples are the Wordnet hierarchy, Google's Product Taxonomy, the Open Directory Project (ODP) etc. Extreme multi-class classification against such taxonomies has been worked on for a relatively long time and several approaches have been explored over the years. The two most common approaches that have been adopted are flat single-step classifiers and hierarchical multi-step classifiers. 

Yu et. al. \cite{Yu_producttitle} explored several word level features in conjunction with linear classifiers like SVMs for classification. Kozareva \cite{Kozareva2015} used several word features (n-grams, LDA, Word2Vec embeddings etc.) followed by linear classifiers. Ha et. al. \cite{Ha_KDD2016} proposed multiple LSTM blocks, one for each unstructured or structured attribute followed by fully-connected layers for classification. Xia et. al. \cite{Xia2017} proposed a variation of CNNs called Attention CNNs applied on Japanese product titles for classification.

Weigend et. al. \cite{Weigend99} used a hierarchical classification scheme with one meta-classifier to determine the broad topic followed by individual topic level classifiers to distinguish nuances within each topic. They also observed that using neural networks in place of standard logistic regression resulted in improved performance. Shen et. al. \cite{Shen12} reformulated this into a two level classification problem ignoring the prior hierarchy and distributing the leaf nodes fairly evenly across top level categories. They used a kNN classifier at the top level followed by individual SVM classifiers for each second level node.

Yundi Li et. al. \cite{YundiLi18} used a Machine Translation approach to generate a root-to-leaf path in a product taxonomy. Gupta et. al. \cite{GuptaKBJ16} trained path-wise, node-wise and depth-wise models (with respect to the taxonomy tree) and trained an ensemble model based on the outputs of these models. Zahavy et. al. \cite{ZahavyMKM16} use decision level fusion networks for multi-modal classification using text and image inputs.

Flat classification models have more parameters per model but perform inference only once per product and hence have a lower latency. We can also batch products together to improve throughput. Hierarchical models on the other hand need to deal with products individually since each product may trace a different path in the taxonomy tree.

All of the above approaches use unstructured product attributes like title, description etc. to perform classification. Ha et. al. \cite{Ha_KDD2016} used a limited set of structured attributes but used an independent LSTM block for each structured attribute which does not scale when each category of products contains different sets of attributes and there are thousands of product attributes overall that need to be considered.

\section{Our Approach}
At Walmart, we experimented with multiple approaches to tackle this problem. Before delving deeper into all our experiments, we will present an overview of the information that a Walmart product contains. These are the attributes that will feed into the Machine Learning models.

\subsection{What does a product contain?}
A product is any commodity that may be sold by a seller. Within our catalog, every product contains a mix of unstructured and structured attributes which describe the product. Examples of unstructured attributes include \textbf{\textit{product name}}, \textbf{\textit{product short description}}, \textbf{\textit{product long description}}, \textbf{\textit{shelf description}}, \textbf{\textit{synopsis}} etc. Examples of structured attributes include \textbf{\textit{screen size}}, \textbf{\textit{color}}, \textbf{\textit{gender}}, \textbf{\textit{fabric material}}, \textbf{\textit{hard drive capacity}} etc. Structured attributes may have a closed list or an open list of values. For example, \textbf{\textit{gender}} has a closed list of values while \textbf{\textit{brand}} is an open list. Every product may also have multiple product images associated with it. These structured and unstructured attributes along with product images can all potentially be used to categorize the product.

\subsection{Hierarchical Classification Approach}
We first experimented with a hierarchical classification scheme similar to Weigend et. al. \cite{Weigend99} using a bag-of-words hash feature on titles and descriptions followed by an entropy maximization based multinomial logistic regression classifier at each node in the taxonomy tree. The training data would vary based on the node the classifier was being trained for but the extracted features and classification scheme remained the same. This approach had several advantages and disadvantages.

\begin{itemize}
    \item Advantages:
    \begin{itemize}
        \item This architecture lent itself to parallelization at training time since each of the individual models could be trained in parallel.
        \item Considering the rate at which the product catalog changes, we could focus specifically on retraining the parts of the model that needed attention rather than retrain the entire model at every iteration.
        \item We could also target specific models for improvement without adversely affecting the performance of the rest of the models in the hierarchy.
    \end{itemize}
    \item Disadvantages:
    \begin{itemize}
        \item Top-1 predictions were very fast but top-k predictions were extremely slow since multiple paths (potentially all paths) along the hierarchy needed to be explored before returning the top-k predictions.
        \item Large fraction of errors were made at the root level and the performance of the overall model was bounded above by the performance of the model at the root.
    \end{itemize}
\end{itemize}

This hierarchical model is easy to deploy and could run efficiently without requiring GPUs. However, the taxonomy hierarchy itself had a significant impact on the performance of the model at the root. For example, product types \textbf{\textit{Athletic Shoes}} and \textbf{\textit{Dance Shoes}} appear under the category \textbf{\textit{Sports \& Outdoor}} while product types \textbf{\textit{Casual \& Dress Shoes}} and \textbf{\textit{Safety Shoes \& Boots}} appear under the category \textbf{\textit{Clothing, Shoes \& Accessories}} even though all 4 product types may be close to each other with respect to product content. Such confusions make it harder for the root node classifier to distinguish between categories at a high level for several types of products. After conducting an analysis, we found that this indeed contributes the largest fraction to the drop in accuracy.

\begin{figure*}[ht!]
  \centering
  \includegraphics[width=\linewidth]{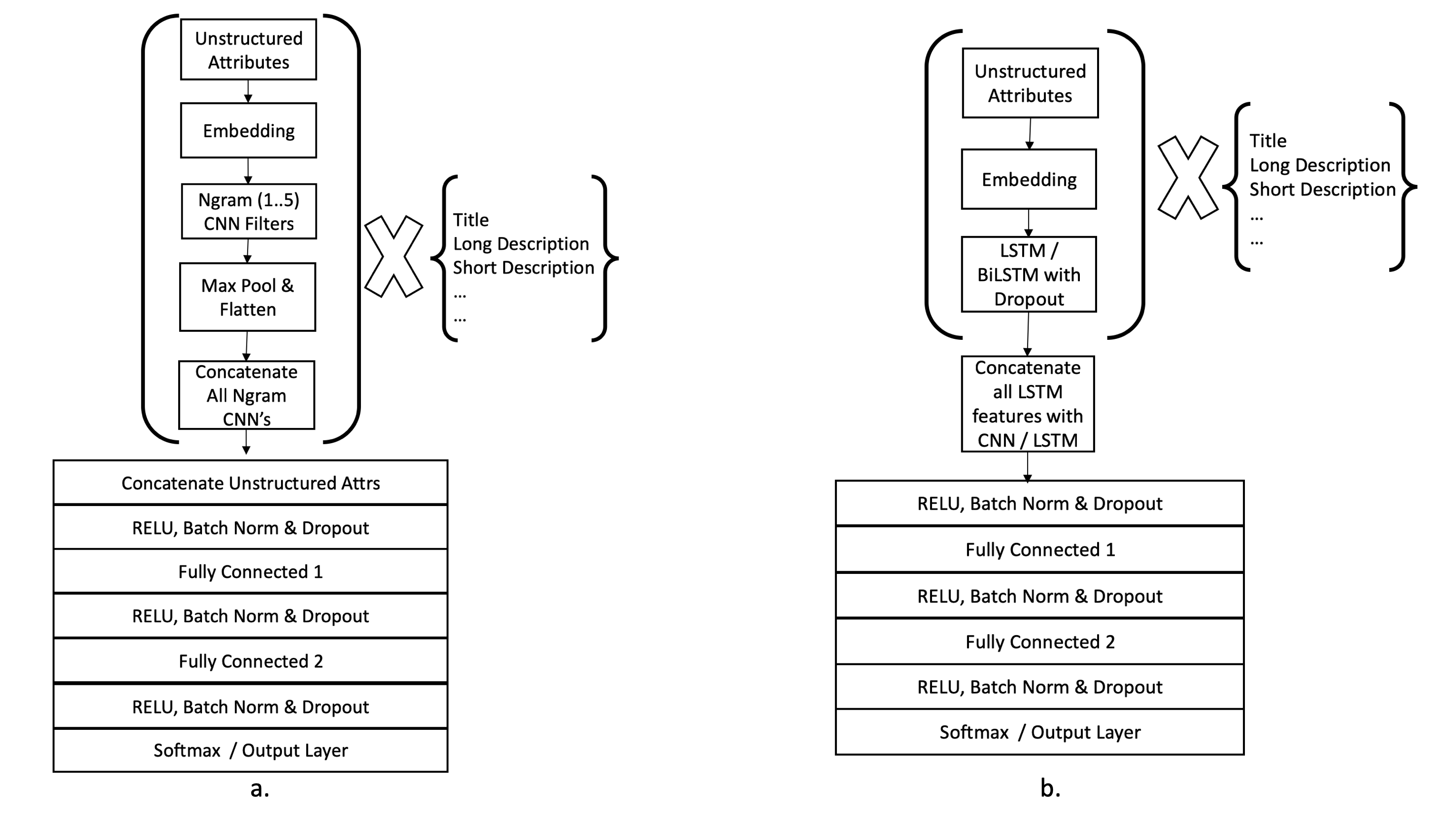}
  \caption{Model architectures used for unstructured attributes. Figure a. shows the Multi-CNN architecture and Figure b. shows the Multi-LSTM architecture.}
  \label{fig:model_arch_unstructured}
\end{figure*}

\subsection{Flat Classification Approach}
Recently, we explored a single-step flat categorization approach using Deep Learning based models. These models seemed to benefit from the large amount of labeled data we had acquired and also demonstrated robustness to random label noise compared to the hierarchical model. In this paper, we describe a Multi-LSTM and a Multi-CNN based approach to perform product categorization. We also present a novel way to use structured attributes that can scale to any number of product attributes and any cardinality in the value space for a particular attribute. This method of using the structured attributes can be added on to any baseline model architecture and should provide a significant boost in classification performance. We observed that it achieves very similar improvements over both the Multi-LSTM and Multi-CNN architectures.

\begin{table*}[!ht]
  \caption{Example product}
  \label{tab:example_product}
  \begin{tabular}{c|c}
    \toprule
    Attribute Name & Attribute Value\\
    \midrule
    product\_name & Rails Womens Plaid Spread Collar Button Down Top\\
    product\_short\_description & This Rails Button Down Top is guaranteed authentic. It's crafted with 100\% Rayon.\\
    assembled\_product\_weight & 0.5 Pounds\\
    color & White\\
    clothing\_size\_type & Regular\\
    clothing\_size\_group & Women\\
    maternity & N\\
    age\_demographic & Women\\
    brand & Rails\\
    fabric\_material & Jersey\\
    clothing\_size & S\\
    style\_sleeve & Long Sleeves\\
    actual\_color & White Navy Sky\\
    country\_of\_origin\_assembly & CN\\
    personalizable & N\\
    \bottomrule
  \end{tabular}
\end{table*}

\subsubsection{\textbf{Generating Word Dictionaries}}\hspace*{\fill} \\
For each unstructured attribute, like \textbf{\textit{product name}} or \textbf{\textit{product short description}}, we go through the entire catalog and gather data for the attribute from every available product to build a word dictionary. We then trim this dictionary to retain a subset of words. The number of words we retain for each attribute depends on the original size of the dictionary and the variety of words found for that particular attribute. We use word embeddings with 200 dimensions for each attribute which are initialized randomly and updated over the course of training. We also experimented with Word2Vec and Glove embeddings with and without updates during the course of training. However, using these embeddings degraded performance marginally and we also observed that the dictionaries associated with these pretrained embeddings did not capture a lot of very important tokens in our catalog. We also experimented with embeddings of size as low as 50 and found that these lower dimensional embeddings had comparable performance to embeddings with 200 dimensions.

\subsubsection{\textbf{Multi-CNN Model}}\hspace*{\fill} \\
In this approach, we use an independent set of convolutional filters of multiple lengths (1,2,3,4,5) on the word embeddings for each unstructured attribute, as shown in Figure \ref{fig:model_arch_unstructured}a. 128 filters of each length are used and each filter is of size \textit{n*200} where \textit{n} represents the filter size. This approach is similar in motivation to Kim \cite{Kim2014}. The multiple length convolutional filters in essence capture n-gram features from pieces of text. The activations generated by each set of convolutional filters are passed through a max pooling layer along the last dimension to retain one activation per filter. These activations for each attribute are then concatenated and passed through multiple fully connected layers followed by a softmax layer at the end for classification. The performance of this approach was very similar to that of the Bidirectional Multi-LSTM model and led to an improvement of almost 20\% over the hierarchical model.

\subsubsection{\textbf{Bidirectional Multi-LSTM Model}}\hspace*{\fill} \\
In this approach, we use one Bidirectional LSTM layer on the embeddings for each unstructured attribute, as shown in Figure \ref{fig:model_arch_unstructured}b. The activations generated by each LSTM layer can either be concatenated directly or put through another bidirectional LSTM following which we add multiple fully-connected layers and softmax at the end for classification. This approach is similar to the one used by Ha et. al. \cite{Ha_KDD2016}, however, we found that using a second level LSTM that takes in the activations of the first level of LSTMs as time steps performs slightly better than just concatenating the activations and using fully connected layers on top. Using this approach led to an improvement of almost 20\% over the hierarchical model.

\subsection{Using Structured Attributes}
Every product has a small set of structured attributes which varies based on the category. Some attributes are common across categories while others are specific to a particular category. For example, \textbf{\textit{assembled\_product\_width}} is an attribute that is relevant to both \textbf{\textit{Cell Phones}} and \textbf{\textit{End Tables}} even though these product types appear under completely different categories. In this case, the value of the attribute may have some useful information regarding the category or product type to which the product could potentially belong. On the other hand, an attribute like \textbf{\textit{diaper\_size}} is only relevant to diaper related product types. In such cases, just the presence of the attribute is a strong indicator about the product type.

\subsubsection{\textbf{Traditional Use of Structured Attributes}}\hspace*{\fill} \\
Guo and Berkhahn \cite{Guo2016} proposed Entity Embeddings for Categorical Variables for the Rossmann Store Sales Kaggle competition. This approach helps avoid feature sparsity and captures semantic relationships between the entities in a euclidean space. This is the latest state of the art method using structured attributes. However, one of the limitations of this method is that separate entity embeddings are needed for each categorical variable which quickly becomes unmanageable when the number of attributes grows to a few thousand. In addition, lots of attributes whose value spaces are open list are not necessarily categorical and hence may need other representations.

\subsubsection{\textbf{Building Attribute Word Dictionaries}}\hspace*{\fill} \\
We build a common word dictionary across all attribute names and values unlike our approach with unstructured attributes where we build a separate dictionary for each attribute. This common dictionary is then trimmed down based on the total number of words in it. It is recommended to explicitly ensure that all the tokens present in the complete set of product attribute names are present in the dictionary after it has been trimmed down. This is accomplished by building a separate dictionary for tokens in the attribute names alone and then merging it with the joint dictionary for attribute names and values. This enables the model to extract richer features from these attribute name value pairs.

\begin{figure*}[ht!]
  \centering
  \includegraphics[width=\linewidth]{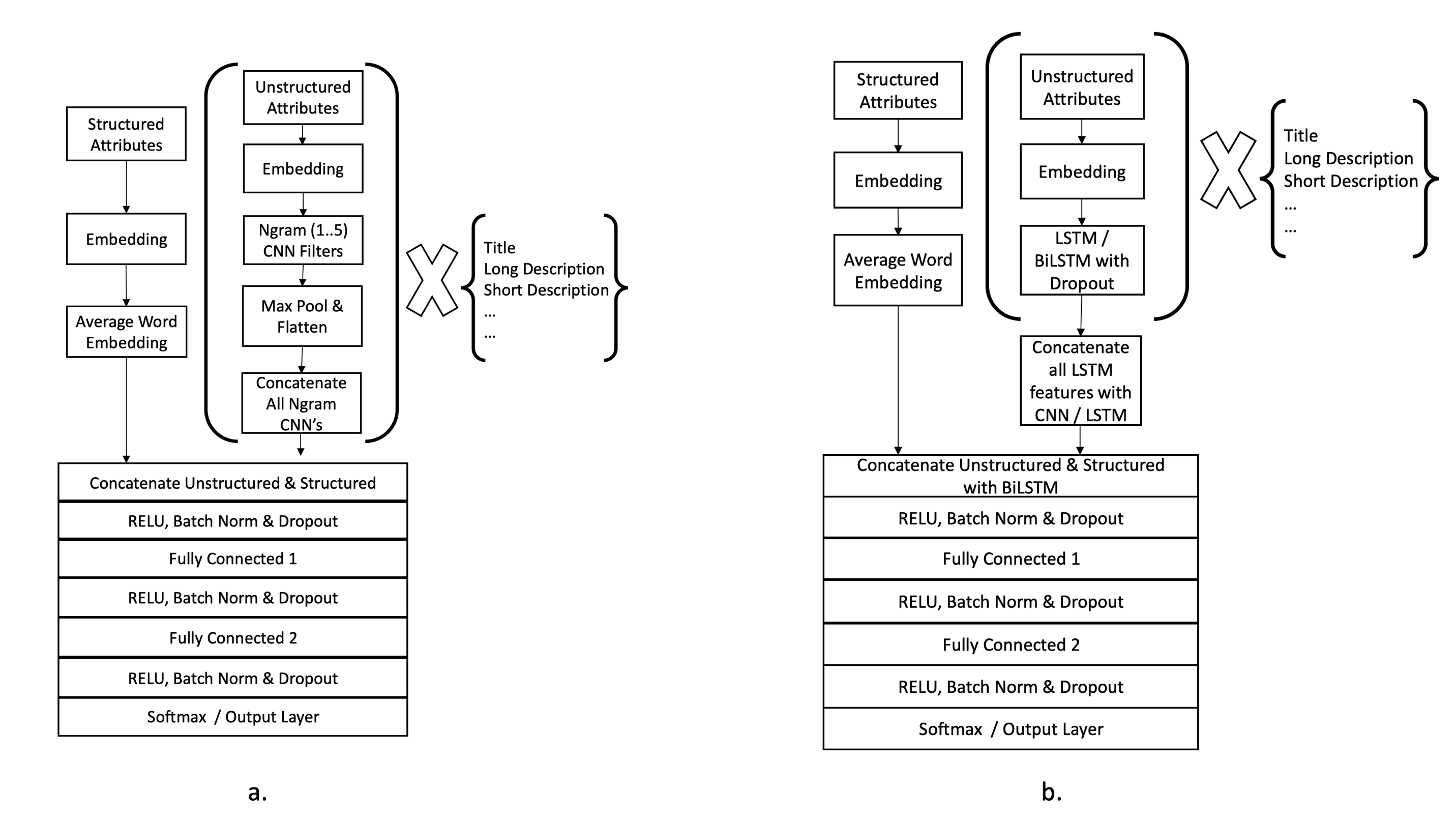}
  \caption{Model architectures used for unstructured and structured attributes using averaged word embeddings . Figure a. shows the Multi-CNN architecture and Figure b. shows the Multi-LSTM architecture.}
  \label{fig:model_arch_avg_embedding}
\end{figure*}

\subsubsection{\textbf{Combining Structured Attributes}}\hspace*{\fill} \\
Once the dictionary is built, we combine all the structured attributes associated with the product in an unstructured fashion. We use both attribute names and values to achieve the best performance. We observed that using just the attribute names improves performance marginally but using both names and values together provides the best results. Attribute names and values are broken down into natural language by stripping out underscores, dashes and other such special delimiters and then combined in the order below.\\

\fbox{\begin{minipage}{24em}
<attr\_name> <attr\_value> <separator> <attr\_name> <attr\_value> <separator> ... <attr\_name> <attr\_value>
\end{minipage}
}\\

This is very similar in format to other unstructured attributes like \textbf{\textit{product\_name}} or \textbf{\textit{product\_long\_description}}. The custom separator token is added so that the convolutional filters have a marker depicting the end of one attribute and the beginning of the next attribute. In our experiments, we found that using the separator token improves classification accuracy over not using it. An example product is shown in Table~\ref{tab:example_product}. The combined structured attribute set for this product is shown below:\\

\fbox{\begin{minipage}{24em}
assembled product weight 0.5 Pounds \_\_sep\_\_ color White \_\_sep\_\_ clothing size type Regular \_\_sep\_\_ maternity N \_\_sep\_\_ clothing size group Women \_\_sep\_\_ age demographic Women \_\_sep\_\_ brand Rails \_\_sep\_\_ fabric material Jersey \_\_sep\_\_ clothing size S \_\_sep\_\_ style sleeve Long Sleeves \_\_sep\_\_ actual color White Navy Sky \_\_sep\_\_ country of origin assembly CN \_\_sep\_\_ personalizable N.
\end{minipage}}\\

\subsubsection{\textbf{Structured Attribute Features using Word Averaging}}
Wieting et. al. \cite{Wieting2016} showed that a simple word averaging method performs well on sentiment classification. We use the same method as a feature extractor and create a joint embedding for all structured attribute names and values in a given product. The joint embedding is simply the average of the word embeddings of the tokens found in the structured attribute names and values. This feature is then concatenated with the features extracted by the initial layers of the Multi-LSTM or the Multi-CNN models from the unstructured attributes. This is considered the baseline approach to incorporate structured attributes in our classification model. We observe that even this simple averaging of embeddings is an extremely useful feature and results in a lift in classification accuracy. The model architecture for this approach is shown in Figure \ref{fig:model_arch_avg_embedding}.

\subsubsection{\textbf{Structured Attribute Features using Convolutional Filters}}
\begin{figure*}[ht!]
  \centering
  \includegraphics[width=\linewidth]{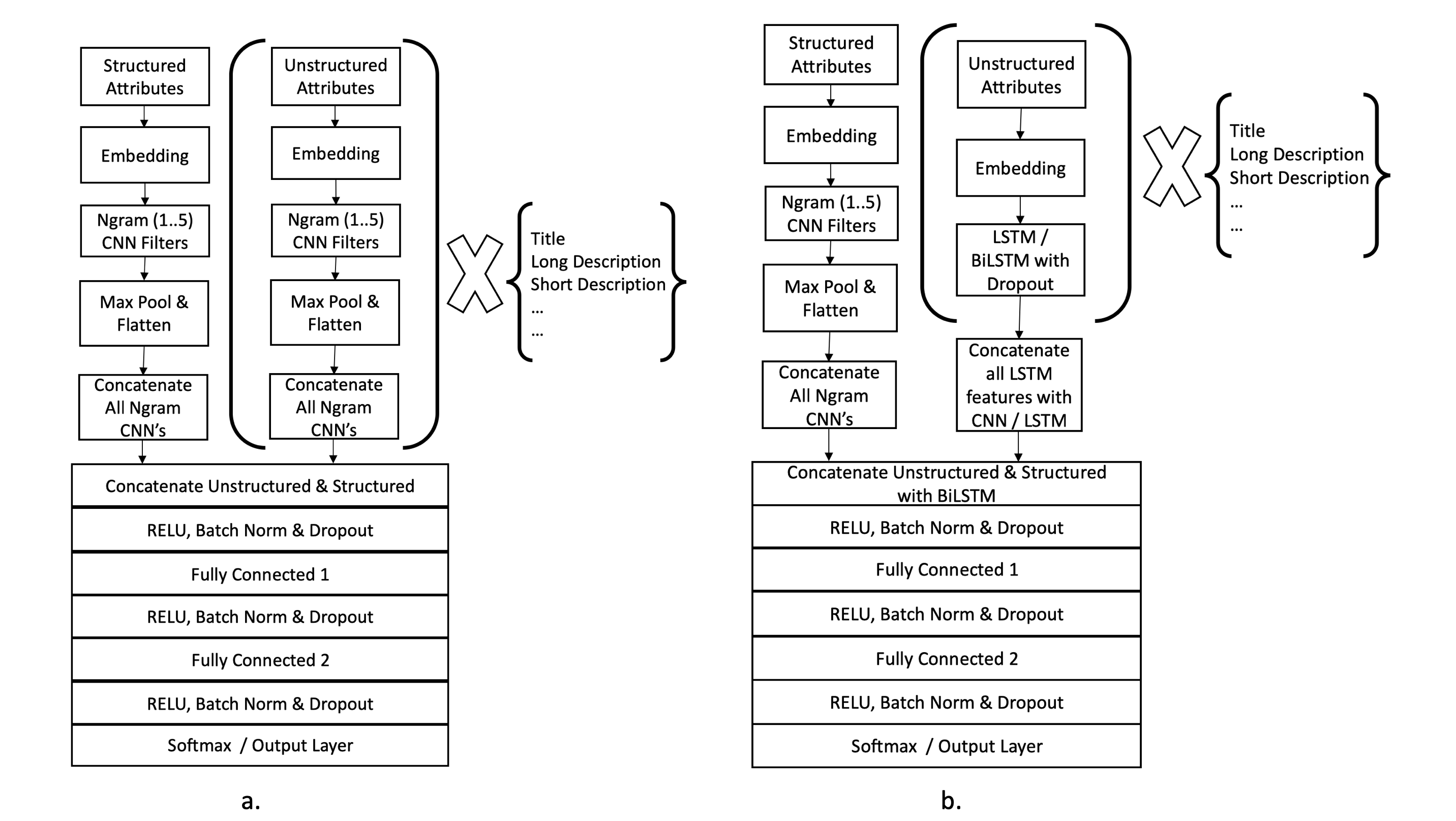}
  \caption{Model architectures used for unstructured and structured attributes. Figure a. shows the Multi-CNN architecture and Figure b. shows the Multi-LSTM architecture.}
  \label{fig:model_arch_structured}
\end{figure*}

\hspace*{\fill} \\
We use the same model architecture as the Multi-CNN model mentioned above to extract features from the combined structured attribute string. These features are concatenated with the features extracted by the Multi-LSTM or the Multi-CNN model. We use convolutional filters of multiple lengths (1,2,3,4,5) that capture features from the pairs of attribute names and values. There are 128 filters for each filter length. We originally experimented with one channel of convolutional filters for attribute names alone and another channel of convolutional filters for both attribute names and values. However, this did not improve performance over just using one channel for both attribute names and values. This is likely because the filters with smaller lengths typically capture features from the attribute names while the filters with larger lengths capture features from the attribute names and values together. Thus both types of features are being captured by the same channel itself. Figure \ref{fig:model_arch_structured} shows the updated model architectures with the block for structured attributes added.

\subsubsection{\textbf{Advantages of Using Structured Attributes in an Unstructured Format}}\hspace*{\fill} \\
There are several advantages to using structured attributes in an unstructured format along with convolutional filters as described above.
\begin{itemize}
    \item Typically structured attributes are used as hand engineered features that are directly concatenated to the final fully connected layers. Our approach removes the need to hand engineer features for each individual attribute since the convolutional filters are able to capture interesting features relevant to the classification problem.
    \item With the current state of the art approaches for using structured attributes as explained in the previous section, there is an inherent sparsity in the feature set. Also, the number of parameters required for the proposed method increases linearly with the number of attributes and the representation format of each additional attribute.
    \item Since we break down the attribute names and values into natural language tokens, this approach should generalize well to new attributes and new values added to existing closed-list or open-list attributes.
    \item By adding the features extracted from structured attributes, we can get better representations for products in general and these representations can be used for a variety of other tasks.
    \item Using word embeddings to represent the tokens present in these attributes also helps capture semantic relationships between multiple attributes or between attributes and their values. Common representations can also be learned for entities identified in these attribute values which can be shared with the other channels for unstructured attributes and potentially for other tasks.
\end{itemize}

However, if the number of structured attributes available is small and limited, traditional approaches may outperform this method.

\section{Experimental Setup}

\subsection{Data}

\begin{figure}[ht!]
  \centering
  \includegraphics[width=\linewidth]{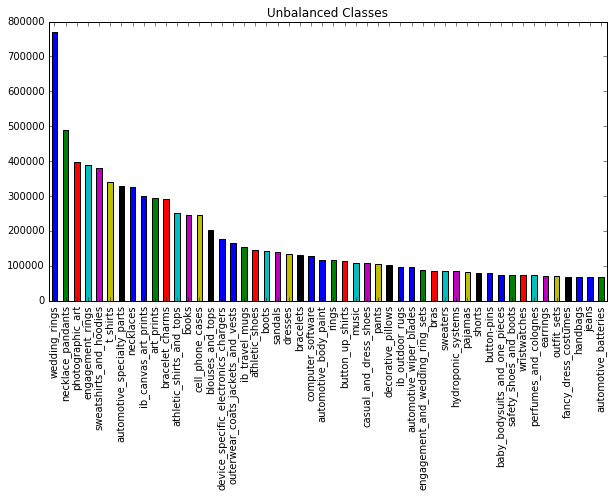}
  \caption{Unbalanced class distribution in the dataset}
  \label{fig:unbalanced_classes}
\end{figure}

Over the years, internal and external crowd sourcing has enabled us to collect large quantities of labeled data for product type categorization. Currently our dataset contains approximately 25 million labels, which is split into 3 subsets (80-10-10) to generate train, validation and test sets. These products represent approximately 6000 leaf product types in our taxonomy. We adopted a bootstrapping approach to collect the external crowdsourced data where we trained a model with the existing data and sent out suggestions to the crowd, who would provide us with feedback. While this approach has helped us collect massive quantities of data, it also has an inherent label bias since the crowd does not have an intimate knowledge of the Walmart taxonomy and is likely to pick a close-enough label if the right label is not presented in the suggestions provided. Our hierarchical model is especially sensitive to this kind of label noise since it uses traditional logistic regression based classifiers. The dataset is also very unevenly distributed for each product type. An example of this uneven distribution is shown in Figure \ref{fig:unbalanced_classes}. This is typical in an eCommerce catalog, where some product types, like \textbf{\textit{T-Shirts}} have several tens of millions of products while others like \textbf{\textit{Zithers}} have less than ten products. We stratify low support classes by repeating samples until each class has at least 200 samples. We also ensure that we perform a stratified split into train, validation and test sets such that the sample distributions across these sets are similar.

\subsection{Preprocessing}
Standard whitespace tokenization is used to tokenize both unstructured and structured attributes. Tokenization and dictionary lookups are not precomputed but performed on the fly during training. Preprocessing is an expensive step however and does slow down training when done on the fly. In our case, since the product information keeps changing frequently, we typically don't store preprocessing results ahead of time before training.

\subsection{Training Schedule}
For both the Multi-LSTM and the Multi-CNN model, we adopt similar training procedures. We use SGD with restarts to train these models as described in Smith \cite{Smith2017CyclicalLR}. We start with one epoch and double the cosine annealing period with each cycle. The models with the lowest validation loss values are usually seen after around 5-6 epochs of training. Even though we also update the randomly initialized word embeddings at training time, the model converges to a loss value within 10\% of the final loss value within one epoch after which it slowly improves and reaches its peak.

\subsection{Dictionary and Embedding Details}
As mentioned above, word embeddings with 200 dimensions were used for each unstructured attribute. The same embedding size was used for structured attributes too. The dictionary used for \textbf{\textit{product\_name}} had 500K tokens, for \textbf{\textit{product\_description}} had 1 million tokens and for \textbf{\textit{structured attributes}} had 100K tokens.

\subsection{Model Details and Training Time}
The hierarchical model has approximately 1.5 billion parameters since it uses one model at each node in the taxonomy tree. The Multi-LSTM model has approximately 180 million trainable parameters while the Multi-CNN model has approximately 330 million trainable parameters. However, we have achieved similar results where both the models were compressed to around 65 million trainable parameters.

The hierarchical model is embarrassingly parallel, but uses CPUs to train and hence takes almost a day to train. The Multi-CNN model with structured attributes takes around 2.5 hours to complete one epoch while the Multi-LSTM model with structured attributes takes around 7 hours. Considering the training time, inference time and overall model performance on the test set, the Multi-CNN model wins over the other two model architectures.

\subsection{Hardware and Software Details}
\begin{itemize}
    \item P100 GPUs used to train the CNN and LSTM models 
    \item LSTM models were trained on Keras while the CNN models were trained on PyTorch
    \item Hierarchical Model trained using scikit-learn and parallelized using Spark
\end{itemize}

\begin{figure}[ht!]
  \centering
  \includegraphics[width=\linewidth]{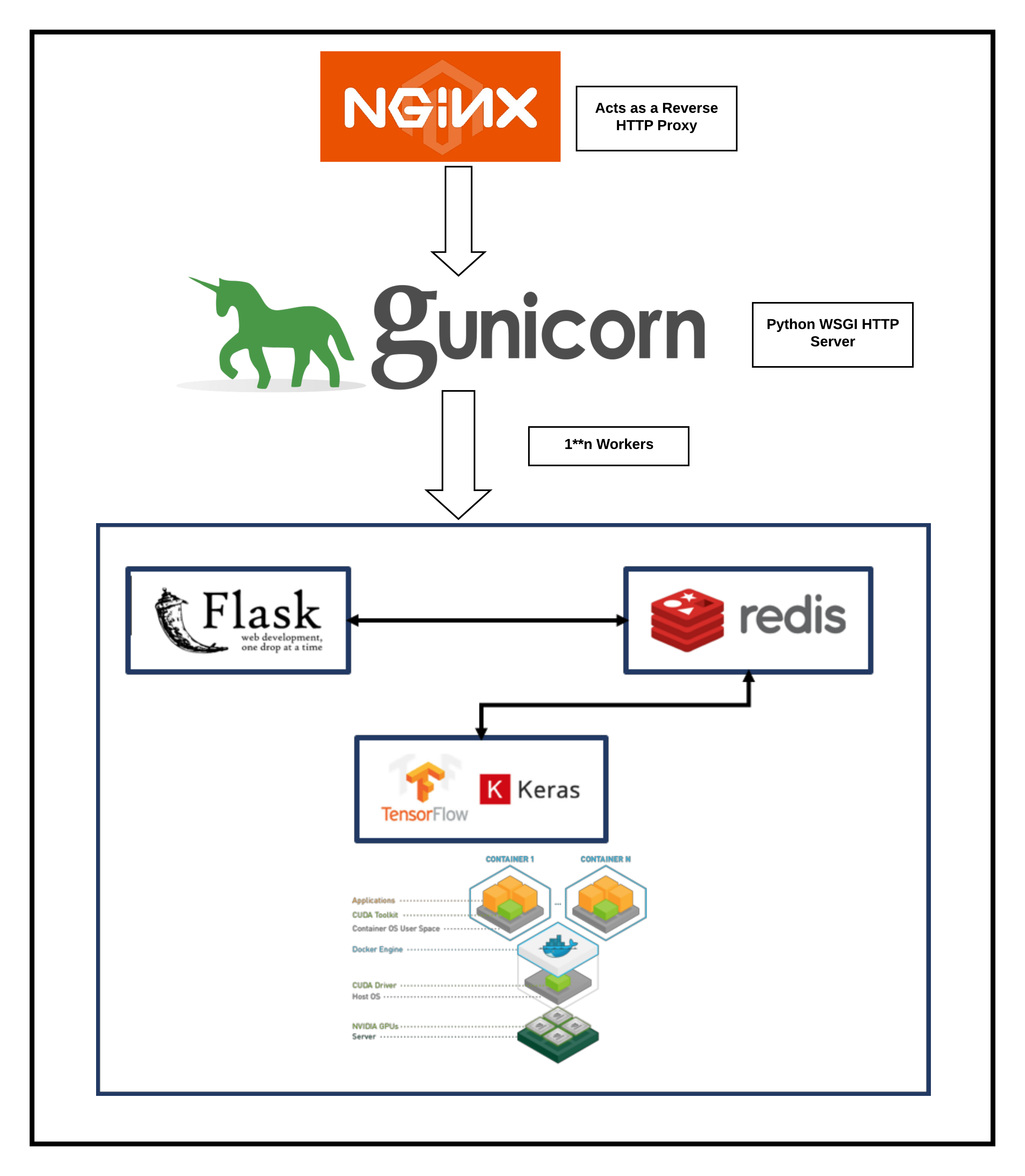}
  \caption{Deployment Architecture}
  \label{fig:deployment_architecture}
\end{figure}

\subsection{Deployment}

\begin{table*}[ht!]
  \caption{Model evaluation results}
  \label{tab:model_comparison}
  \begin{tabular}{c|c|c|c}
    \toprule
    Model architecture & Top-1 acc. & Top-2 acc. & Top-3 acc.\\
    \midrule
    Hierarchical & 70\% & - & -\\
    Multi-LSTM & 89.9\% & 94.79\% & 96.42\%\\
    Multi-LSTM + Word Embedding Avg. & 91.4\% & 95.95\% & 97.18\%\\
    Multi-LSTM + Struct. Attr. & 92.28\% & 96.29\% & 97.38\%\\
    Multi-CNN & 89.45\% & 94.93\% & 96.5\%\\
    Multi-CNN + Word Embedding Avg. & 89.97\% & 95.31\% & 96.77\%\\
    Multi-CNN + Struct. Attr. w/o Separator & 91.08\% & 95.87\% & 97.18\%\\
    Multi-CNN + Struct. Attr. & 92.15\% & 96.36\% & 97.51\%\\
    \bottomrule
  \end{tabular}
\end{table*}

Our Machine Learning models are typically deployed as Micro Services to which other clients can send requests individually over HTTP. We have carefully designed our system with ideas borrowed from Rosebrock \cite{Rosebrock2018} to maximize overall throughput and minimize latency as shown in Figure \ref{fig:deployment_architecture}. Since GPUs are suited for batch work loads, we microbatch multiple requests from different clients by adding a Redis buffering layer. These queued requests are sent as a minibatch to utilize the GPU effectively. A configurable poll interval (currently 0.3s) and a batch size of 1024 were chosen based on latency vs throughput trade-offs. In our load tests, we observed a throughtput of ~750 rps on a single P100 GPU with 6 CPU Cores.

\begin{table}[ht!]
  \caption{Top-5 product types that benefit from using structured attributes with support >= 100 products}
  \label{tab:top_5_pts}
  \begin{tabular}{c|c}
    \toprule
    Product Type & $\Delta$ f1 Score\\
    \midrule
    Tank Tops & 0.365\\
    Chemistry Experiment Kits & 0.317\\
    Office Boxes & 0.257\\
    Outdoor Flags \& Banners & 0.249\\
    Shape Sorting Toys & 0.241\\
    \bottomrule
  \end{tabular}
\end{table}

\section{Results}

We have observed that attaching the structured attributes block (word embeddings followed by convolutional filter activations) to either model architecture improves overall accuracy by approximately 2.7\% on the evaluation set as shown in Table~\ref{tab:model_comparison}. Considering that we have 6000 classes to classify against, this is a very significant increase in accuracy. This also shows that regardless of the model architecture, using this block for structured attributes improves overall accuracy. The model architectures we use, Multi-LSTM or Multi-CNN also lend themselves to variable sized inputs and hence can handle arbitrarily large unstructured attribute values. Upon performing an analysis of the top product types (with a support of over 100 products) in the evaluation set that benefit from using structured attributes, we found that \textbf{\textit{Tank Tops}}, had the highest lift in f1 score. This is due to the presence of multiple attributes like \textbf{\textit{sleeve\_style}} and \textbf{\textit{clothing\_top\_style}} whose values indicate that the product is a \textbf{\textit{Tank Top}}. Similarly, another product type that sees a significant lift in f1 score is \textbf{\textit{Office Boxes}}. This is again because of the presence of the attribute \textbf{\textit{office\_box\_type}} whose presence itself indicates that the product is an \textbf{\textit{Office Box}}. From the above examples, we see that there are two useful features extracted from the structured attributes.
\begin{itemize}
    \item The presence or absence of a particular attribute
    \item The value of a particular attribute
\end{itemize}
The convolutional filters are able to extract both these features efficiently from the structured text. Table \ref{tab:top_5_pts} shows the top 5 product types ordered by the lift in f1 score using this approach.

\section{Conclusions}
Several insights, observations and conclusions were derived from this large-scale experiment some of which are mentioned below:

\begin{itemize}
    \item \textbf{Training size}: For such extreme classification problems, we observe that having a large dataset with a good variety of samples to train significantly improves overall accuracy. Stratifying the dataset to improve representation for low support classes has also been shown to improve performance.
    \item \textbf{Choice of model}: In the presence of such large amounts of data, Deep Learning models significantly outperform traditional Machine Learning models. In this case, both Multi-LSTMs and Multi-CNNs perform equally well. However, the Multi-CNNs have the advantage of being more parallelizable and hence will be faster both at training and inference time.
    \item \textbf{Taxonomy structure}: Having the right taxonomy structure significantly improves the performance of classification models. Logical and mutually exclusive divisions of product types into categories helps improve classification performance significantly, especially using hierarchical models. Having more specific product types than very general ones is also recommended to aid better classification.
    \item \textbf{Word embeddings}: Using separate embedding spaces for each unstructured attribute seemed to perform better than using a common embedding space for all attributes. This is likely because the semantic relationships between words present in each attribute may vary.
    \item \textbf{Word embedding size}: Most of our experiments were performed with 200 dimensional word embeddings. However, embeddings as small as 50  dimensions also yielded similar results. Also, trimming the word dictionaries for each attribute seemed to act as an implicit regularizer and helped performance in some cases.
    \item \textbf{Using a separator}: While concatenating the structured attribute names and values together, it is beneficial to use a separator token between two attributes. This helps the model not to relate the values of one attribute with a different attribute name.
\end{itemize}

%
\bibliographystyle{ACM-Reference-Format}
\bibliography{references}

\end{document}